\documentclass[prb,aps,twocolumn,amsmath,amssymb,showpacs]{revtex4}
\usepackage{graphicx}
\usepackage{psfrag}
\usepackage{color}
\usepackage{subfigure}
\usepackage{amsmath}
\definecolor{dred}{rgb}{0.7,0.0,0.0}
\allowdisplaybreaks 
\usepackage{array}

\newcommand{\e}{{\rm e}}

\begin{document}

%
%


\title{Magnetic properties and Mott transition of the Hubbard model for weakly coupled chains on the 
anisotropic triangular lattice}
\author{A. Yamada}
\affiliation{Department of Physics, Chiba University, Chiba 263-8522, Japan}

\date{\today}

\begin{abstract}
We investigate the magnetic properties and Mott transition in the Hubbard model for weakly coupled chains 
on the anisotropic triangular lattice. 
Taking into account 120$^\circ$ N\'eel, and collinear orderings, 
the magnetic phase diagram is studied at zero temperature and half-filling by the variational cluster approximation. 
We found that when the on-site Coulomb repulsion $U$ is relatively large, nonmagnetic insulator, 
which is a candidate of the spin liquid state, is realized for wide range of the interchain hopping from 
quasi two-dimensional to almost one-dimensional regime. 
When the interchain hopping is relatively large, this nonmagnetic insulator becomes magnetic states as $U$ decreases.
For rather small interchain hopping, it changes to the paramagnetic metal, 
thus purely paramagnetic metal-insulator transition (Mott transition) takes place. 
Implications of our results for the Cs$_2$CuBr$_4$ and Cs$_2$CuCl$_4$ are discussed. 

\end{abstract}
 
 
\maketitle

%
%

\section{Introduction}

\begin{figure}
\includegraphics[width=0.47\textwidth,bb= 11 244 584 521]{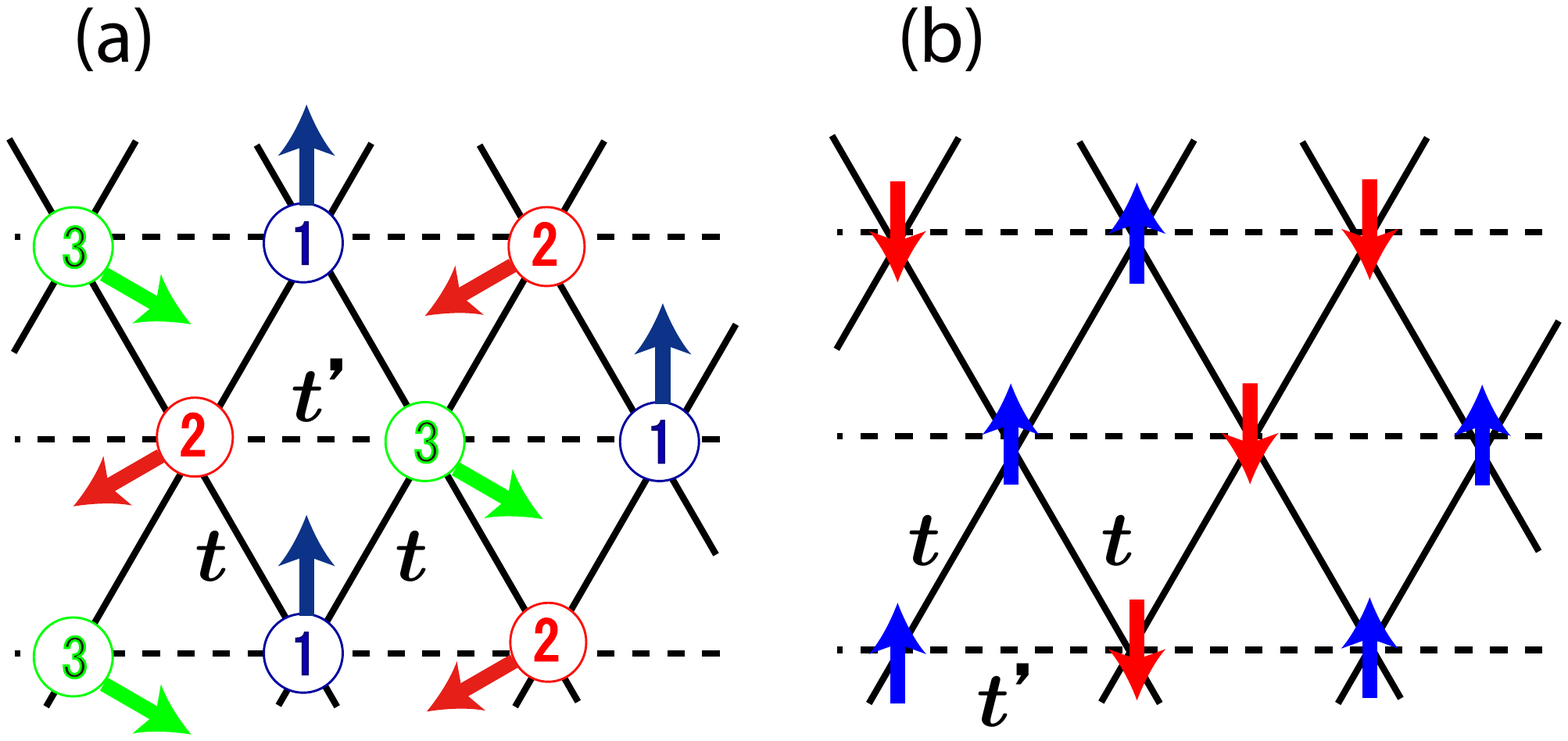}
\caption{(Color online)
The magnetic orderings 
(a) 120$^\circ$ N\'eel (spiral) and 
(b) collinear (AFC) 
on the anisotropic triangular lattice with the hopping parameters $t$ and $t'$.
\label{fig:spin-config}\\[-1.5em]}
\end{figure}

The interplay of electron correlations, low dimensionality, and frustrated magnetic interactions lead to a rich variety of 
phenomena. 
For example, the organic charge-transfer salts $\kappa$-(BEDT-TTF)$_2\mathrm{X}$~\cite{lefebvre00,shimizu03,kanoda3,manna} 
exhibit various phases, e.g., anti-ferromagnetic metal, superconductivity, and quantum spin liquid. 
These compounds are well approximated by the half-filled Hubbard model on the anisotropic triangular lattice described by 
two hopping parameters $t$ and $t'$ in different spatial directions (see Fig.~\ref{fig:spin-config}) 
since (BEDT-TTF)$^+_2$ dimers lie on a quasi-two-dimensional triangular lattice, and provide us useful 
information on the role of the frustration and strong electron correlations in two dimensional systems. 

In addition to the organic charge transfer salts, the material Cs$_2$CuCl$_{4-x}$Br$_x$ is 
also approximated by the half-filled Hubbard model because the (magnetic) copper atoms carrying a spin of 1/2 
form a weakly coupled triangular lattice and shows a variety of magnetic properties.\cite{cong2011} 
In fact one of the end-member of these compounds Cs$_2$CuBr$_4$ exhibits the spiral order,\cite{ono2004} 
while the other end-member Cs$_2$CuCl$_4$, 
which is more one-dimensional compared to Cs$_2$CuBr$_4$,\cite{coldea2001,foyevtsova2011,ono2005,Zheng2005,zvyagin2014} 
becomes two-dimensional spin liquid below $T \leq 2.65$~K until the three dimensionality becomes relevant and 
three-dimensional magnetic order appears at $T = 0.62$~K.\cite{coldea2001} 
Unlike the organic charge-transfer salts which are two-dimensional materials, 
the compound Cs$_2$CuCl$_{4-x}$Br$_x$ corresponds to the weakly coupled (one-dimensional) chains on the triangular lattice 
with $1.2 \lesssim t'/t \lesssim 2.0$,\cite{coldea2001,foyevtsova2011,ono2005,Zheng2005,zvyagin2014} 
and provides us a possibility to investigate the effects of the dimensionality and 
frustration from quasi two-dimensional to quasi one-dimensional region. 
\begin{figure}
\includegraphics[width=0.47\textwidth,bb= 81 267 503 416]{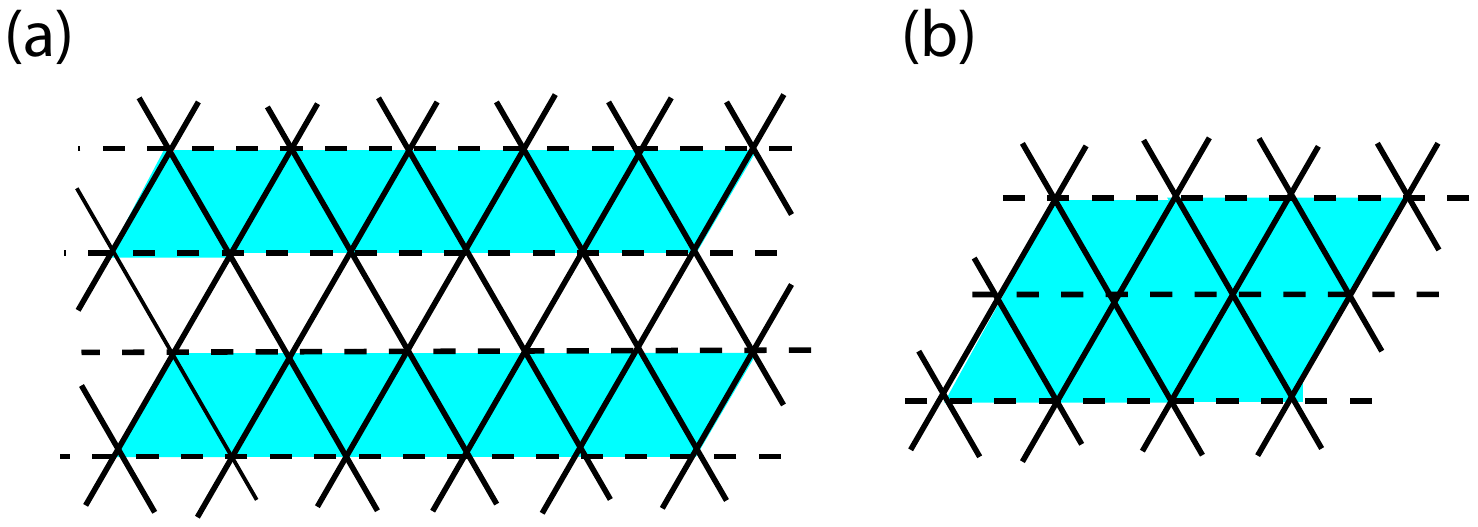}\\
\caption{
(Color online)
(a) The 2$\times$6 and (b) 3$\times$4 clusters (shaded clusters) used in VCA. 
In the analysis of the AFC, the two 2$\times$6 clusters are combined as in (a) 
to cover the infinite lattice consistently in the presence of the AFC ordering in the Bravais sense. 
\label{fig:cluster}
}
\end{figure}

In this paper, motivated by the experiments on the family of compound Cs$_2$CuCl$_{4-x}$Br$_x$, 
we investigate the Hubbard model on the anisotropic triangular lattice in the region of the weakly coupled chains. 
Taking into account 120$^\circ$ N\'eel and collinear orderings (see Fig.~\ref{fig:spin-config}), 
which are referred to as spiral and AFC hereafter, we analyze the magnetic properties and Mott transition 
at zero temperature and half-filling 
by the variational cluster approximation (VCA),\cite{Senechal00,Potthoff:2003-1,Potthoff:2003} 
which is based on a rigorous variational principle and exactly takes into account the short-range correlations 
on the reference cluster used in the analysis. 
When the system evolves towards weakly coupled one-dimensional chains, 
the correlations along the intrachain direction (dotted lines in  Fig.~\ref{fig:spin-config}) become important, 
therefore we use the 2$\times$6 and 3$\times$4 clusters in Fig.~\ref{fig:cluster} as our reference clusters, which contain 
two 6-site chains and three 4-site chains, respectively. 

We found that when the on-site Coulomb repulsion $U$ is relatively large, nonmagnetic insulator, 
which is a candidate of the spin liquid state, 
is realized for wide range of the interchain hopping $1.2 \lesssim t'/t$. 
As $U$ decreases, this nonmagnetic insulator becomes magnetic states for $1.2 \lesssim t'/t \lesssim 1.6$, 
and for more one-dimensional region $1.65 \lesssim t'/t$, 
it changes to the paramagnetic metal, thus purely paramagnetic metal-insulator transition (Mott transition) takes place. 
The implications of our results on the materials Cs$_2$CuCl$_{4-x}$Br$_x$ are discussed. 

Recently a related study by the variational Monte Carlo (VMC)\cite{tocchio2} has reported. 
Comparing our VCA results with those of VMC,\cite{tocchio2} 
the main difference is that the magnetic state is realized between the paramagnetic metal and nonmagnetic insulator 
for whole the range of the parameter space $1 \le t'/t \lesssim 3.3$ studied by VMC,\cite{tocchio2} 
while magnetic state is not realized for $1.65 \lesssim t'/t$ in our VCA analysis.

%
%
%
%

\section{Variational cluster approximation}

The Hubbard model on the anisotropic triangular lattice is described by the Hamiltonian 
\begin{align}
H =& -\sum_{i,j,\sigma} t_{ij}c_{i\sigma }^\dag c_{j\sigma}
+ U \sum_{i} n_{i\uparrow} n_{i\downarrow} - \mu \sum_{i,\sigma} n_{i\sigma},
\label{eqn:hm}
\end{align}
where $c_{i\sigma}$ ($c_{i\sigma }^\dag$) denotes the annihilation (creation) operator for an electron at site $i$ 
with spin $\sigma$, $n_{i\sigma}=c_{i\sigma}^\dag c_{i\sigma}$, 
$t_{ij}=t$ for the solid lines and $t_{ij}=t'$ for the dashed lines in Fig.~\ref{fig:spin-config}, $U$ is 
the on-site Coulomb repulsion, and $\mu$ is the chemical potential, which we include into the Hamiltonian for the later 
convenience. The energy unit is set as $t=1$ hereafter. 

We use VCA\cite{Senechal00,Potthoff:2003-1,Potthoff:2003} in our analysis. This approach 
is an extension of the cluster perturbation theory\cite{Senechal00} based on the 
self-energy-functional approach,\cite{Potthoff:2003} and 
uses the rigorous variational principle 
$\delta \Omega _{\mathbf{t}}[\Sigma ]/\delta \Sigma =0$ for the thermodynamic grand-potential 
$\Omega _{\mathbf{t}}$ expressed as a functional of the self-energy $\Sigma $ by
\begin{equation}
\Omega _{\mathbf{t}}[\Sigma ]=F[\Sigma ]+\mathrm{Tr}\ln(-(G_0^{-1}-\Sigma )^{-1}).
\label{eqn:omega}
\end{equation}%
In Eq. (\ref{eqn:omega}), the index $\mathbf{t}$ denotes the explicit dependence of 
$\Omega _{\mathbf{t}}$ on all the one-body operators in $H$, $F[\Sigma ]$ is the Legendre 
transform of the Luttinger-Ward functional,\cite{lw} and $G_0$ is the non-interacting Green's function of $H$. 
The variational principle $\delta \Omega _{\mathbf{t}}[\Sigma ]/\delta \Sigma =0$ is equivalent to the Dyson's equation and 
$\Omega _{\mathbf{t}}[\Sigma ]$ gives the exact grand potential for the self-energy of $H$ 
which satisfies Dyson's equation. 

All Hamiltonians with the same interaction part have the same functional form of $F[\Sigma ]$, and 
using that property we evaluate $F[\Sigma ]$ for the self-energy of a simpler Hamiltonian $H'$, though 
the space of the self-energies where $F[\Sigma ]$ is evaluated is now restricted to that of $H'$. 
To construct $H'$, we first divide the infinite lattice into identical clusters, 
referred to as reference cluster hereafter, which tile the original infinite lattice by 
removing the hopping parameters between them, and take the Hubbard Hamiltonian on these 
non-interacting identical clusters as $H'$. 
In this construction, $H'$ differs from $H$ only by the hopping terms and $H'$ and $H$ share the same interaction part. 
For $H'$ the grand potential is expressed as a functional of $\Sigma$ as
\begin{equation}
\Omega' _{\mathbf{t'}}[\Sigma ]=F[\Sigma ]+\mathrm{Tr}\ln(-(G'_0{}^{-1}-\Sigma )^{-1}),
\label{eqn:omegaprime}
\end{equation}%
where $G'_0$ is the non-interacting Green's function of $H'$ and $\mathbf{t}'$ denotes all the one-body operators in $H'$. 
In Eqs. (\ref{eqn:omega}) and (\ref{eqn:omegaprime}), $F[\Sigma ]$ is the same for a given $\Sigma$ since the interaction 
part is the same for $H$ and $H'$, therefore subtracting Eq. (\ref{eqn:omegaprime}) from Eq. (\ref{eqn:omega}), 
we obtain a functional relation between $\Omega _{\mathbf{t}}[\Sigma ]$ and $\Omega' _{\mathbf{t'}}[\Sigma ]$ as 
\begin{eqnarray}
\Omega _{\mathbf{t}}[\Sigma ]= \Omega' _{\mathbf{t'}}[\Sigma ] &+& \mathrm{Tr}\ln(-(G_0^{-1}-\Sigma )^{-1})
\nonumber \\
&-&\mathrm{Tr}\ln(-(G'_0{}^{-1}-\Sigma )^{-1}).
\label{eqn:omega3}
\end{eqnarray}%
In Eq. (\ref{eqn:omega3}) $\Omega' _{\mathbf{t'}}[\Sigma ]$ and $\Sigma$ are exactly computed for $H'$ by exactly solving it, 
thus $\Omega _{\mathbf{t}}[\Sigma ]$ is evaluated for the exact self-energy of $H'$. 
In this case, $\Omega _{\mathbf{t}}[\Sigma ]$ is a function of $\mathbf{t}'$ expressed as 
\begin{equation}
\Omega _{\mathbf{t}}(\mathbf{t}')=\Omega' _{\mathbf{t'}} - \int_C{\frac{%
d\omega }{2\pi }} \e^{ \delta \omega} \sum_{\mathbf{K}}\ln \det \left(
1+(G_0^{-1}\kern-0.2em -G_0'{}^{-1})G'\right),
\nonumber
\end{equation}%
where the functional trace has become an integral over the diagonal variables 
(frequency and super-lattice wave vectors) of the logarithm of the determinant over intra-cluster indices, and 
the frequency integral is carried along the imaginary axis with $\delta \rightarrow + 0$. 
The variational principle $\delta \Omega _{\mathbf{t}}[\Sigma ]/\delta \Sigma =0$ becomes the stationary condition 
$\delta \Omega _{\mathbf{t}}(\mathbf{t}') /\delta \mathbf{t}' = 0$, and exact self-energy of $H'$ at the stationary point, 
denoted as $\Sigma^{*}$, are the approximate self-energy of $H$ in VCA. Physical quantities are computed with the 
self-energy $\Sigma^{*}$. 
In VCA, the restriction of the space of the self-energies $\Sigma$ into that of $H'$ 
is the only approximation, and the short-range correlations within the reference cluster are exactly 
taken into account by exactly solving $H'$. 
We analyze a possible symmetry breaking by including the 
corresponding Weiss field in $H'$ which is determined by minimizing the grand-potential $\Omega_{\mathbf{t}}$ 
with respect to the parameters involved in the Weiss field.

In our analysis, we take the 2$\times$6 and 3$\times$4 clusters in Fig.~\ref{fig:cluster} as the reference clusters. 
The cluster shape dependence of the results is a measure of the finite size effects of our analysis. 
To study the magnetic orderings spiral and AFC, the Weiss field 
\begin{eqnarray}
H_{\rm AF}&=& h_{\rm M}\sum_{i} {\bf e}_{a_i} \cdot {\bf S}_i
\label{eqn:weiss}
\end{eqnarray}
with the spin operator ${\bf S}_i = c_{i\alpha}^\dag \sigma_{\alpha\beta}c_{i\beta}$ is included into $H'$, 
where the index $a$ specifies the site in the unit cell in the sub-lattice 
formalism, and $a=1,2,3$ for spiral and $a=1,2$ for AFC.
The unit vectors ${\bf e}_{1,2,3}$ are oriented at 120$^\circ$ of each other for spiral, 
and ${\bf e}_{1}= -{\bf e}_{2} $ for AFC according to this spin orderings (see Fig.~\ref{fig:spin-config}). 
In our analysis the pitch angle of spiral order is fixed to be 120$^\circ$. 
In VCA, we can not study magnetic orderings whose modulation period does not fit into the reference cluster. 

In the analysis of the AFC, we combined the two 2$\times$6 clusters depicted in Fig.~\ref{fig:cluster} (a) 
to cover the infinite lattice consistently in the Bravais sense in the presence of the AFC ordering. 
In this case the Green's function $G'$ of the combined cluster is given by 
\begin{eqnarray}
G'^{-1}&=& \sum_{i} {G'}_i^{-1} + \tilde{t}
\label{eqn:combine}
\end{eqnarray}
where ${G'}_i$ is the exact Green's function on each 2$\times$6 cluster (the site and spin indices suppressed) 
and $\tilde{t}$ is the hopping matrix linking the two 2$\times$6 clusters. 
Even when the two 2$\times$6 clusters are combined as in Fig.~\ref{fig:cluster} (a), 
the Hamiltonian on the 2$\times$6 cluster is exactly diagonalized, therefore 
the correlations within the 2$\times$6 clusters are exactly taken into account. 

In our analysis, we take the Weiss field parameter $h_{\rm M}$ and the cluster chemical potential $\mu'$ in $H'$ 
as the variational parameters, where 
$\mu'$ should be included for the thermodynamic consistency,\cite{aichhorn} 
and search the stationary point of $\Omega(\mu', h_{\rm M})$, which we denote as the grand-potential per site. 
During the search, the chemical potential of the system $\mu$ is also adjusted so that the electron 
density $n$ is equal to 1 within 0.1\%. 
In general, a stationary solution with $h_{\rm M} \neq 0$ corresponding to the magnetically ordered state and 
that with $h_{\rm M} = 0$ corresponding to the nonmagnetic state 
are obtained, and the energies per site $E=\Omega+\mu n$ are compared 
for spiral, AFC, and nonmagnetic state to determine the ground state. 
The density of state per site 
\begin{eqnarray}
D(\omega)= \lim_{\eta \rightarrow 0}  \int
{\frac{%
d^2 k }{(2\pi)^2 }}\frac{1}{n_c}\sum_{\sigma, a=1}^{n_c}\{ -\frac{1}{\pi} \mathrm{Im}G_{a\sigma}(k, \omega+i\eta) \}
\label{eqn:dos}
\end{eqnarray}
is also calculated to examine the insulating gap, where 
$n_c$ is the number of the sites in the unit cell in the sense of the sub-lattice formalism 
($n_c = 3$ for spiral, $n_c = 2$ for AFC and $n_c = 1$ for nonmagnetic state), 
and the $k$ integration is over the corresponding Brillouin zone. 
In Eq. (\ref{eqn:dos}), we evaluate $\eta \rightarrow 0$ limit using the standard extrapolation method 
by calculating $D(\omega)$ for $\eta =0.1$, $0.05$, and $0.025$. 
The numerical error after this extrapolation is of order $10^{-3}$, 
so the gap is identified as the region of $\omega$ around $\omega \simeq 0 $ 
where the extrapolated $D(\omega)$ is less than $10^{-2}$. 
We also compute the magnetic order parameter per site 
\begin{eqnarray}
M&=& \frac{1}{n_c}\sum_{a=1}^{n_c}  {\bf e}_a \cdot \langle {\bf S}_a  \rangle\nonumber
\end{eqnarray}
where $ \langle {\bf S}_a \rangle $ is the expectation value of ${\bf S}_a$.
\begin{figure}
\includegraphics[width=0.48\textwidth,bb= 167 107 501 563]{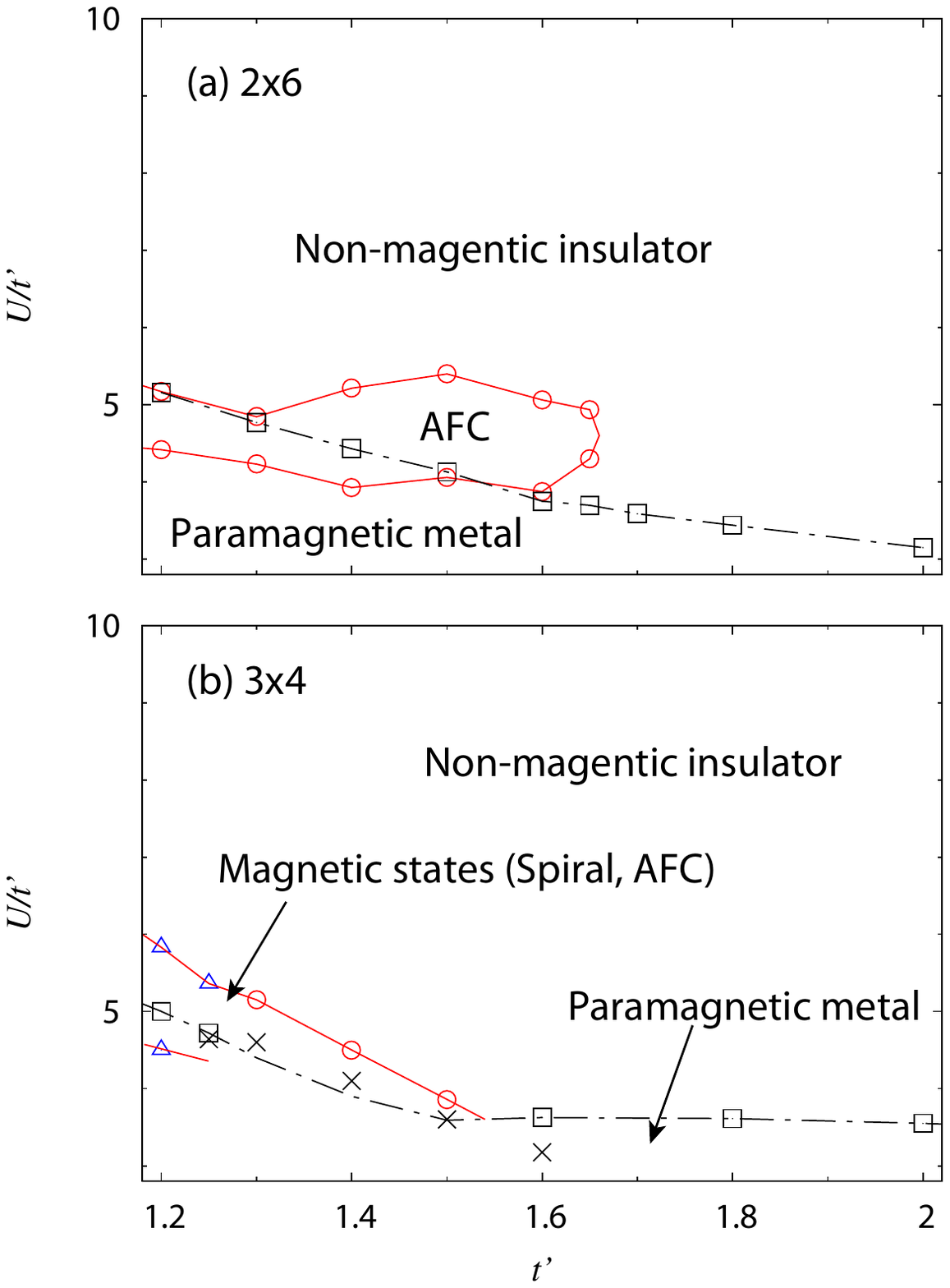}\\[-0.5em]
\caption{
(Color online)
Phase diagrams of the Hubbard model on the anisotropic triangular lattice
at zero temperature and half-filling computed as functions of $t'$ and $U$ by VCA on 
(a) 2$\times$6 and (b) 3$\times$4 clusters. Lines are guides to the eye.
The triangles and circles correspond to spiral and AFC transition points, and squares are the Mott transition points 
obtained assuming that no magnetic order is allowed. 
In (b) the crosses denote the points below which the paramagnetic ground states on the cluster become spin triplet, 
which will be artifacts due to the small system size. 
The Mott gap closes continuously at the squares, 
and it closes discontinuously at the crosses for $ 1.25 \leq t' \leq 1.5$ in (b).
\label{fig:phase-diagram}
}
\end{figure}

\section{Phase diagram}

Fig.~\ref{fig:phase-diagram} shows the phase diagrams of the Hubbard model on the anisotropic triangular lattice
at zero temperature and half-filling as functions of $t'$ and $U$ obtained by VCA on the 
(a) 2$\times$6 and (b) 3$\times$4 clusters. Lines are guides to the eye.
The triangles and circles correspond to spiral and AFC transition points, and 
squares are the Mott transition points computed assuming that no magnetic order is allowed. 
We have investigated the magnetic properties up to $U \simeq 30$ and could not find the magnetically ordered states. 
In (b) the crosses denote the points below which the spin-triplet ground states appears in the exact diagonalization of 
the cluster Hamiltonian for the paramagnetic solutions ($h_{\rm M} = 0$), 
which will be artifacts due to the small system size intrinsic to the 3$\times$4 cluster. 
Therefore the magnetic properties are not analyzed below these points. 
In these figures, the Mott gap closes continuously at the points denoted by squares 
and it closes discontinuously at the crosses in (b) for $ 1.25 \leq t' \leq 1.5$. 
As is stated, the metal-insulator transitions observed at the crosses will be artifacts due to the 
small system size. Energetically disfavored magnetic solutions are not obtained outside the magnetic regions. 

Comparing Fig.~\ref{fig:phase-diagram} (a) and (b), even though there remains some cluster shape dependence, the 
general features are the same. 
When $U$ is relatively large ($ 5\sim 6 \lesssim U/t'$), nonmagnetic insulator is realized for $1.2 \lesssim t' \lesssim 2.0$. 
As $U$ decreases, magnetic states appears for $1.2 \lesssim t' \lesssim 1.6$. 
For $1.6 \lesssim t' \lesssim 2.0$, the non-magnetic insulator changes to paramagnetic metal, thus 
purely paramagnetic metal-insulator transition takes place. 
The transition from nonmagnetic insulator to magnetic states is of the second order 
since the energetically disfavored magnetic solutions are not obtained above the magnetic phase, and 
the Mott transition is of the second order at the transition points denoted by the squares since the Mott gap closes 
continuously and energetically disfavored nonmagnetic states are not obtained near the transition points. 

As for the nature of the magnetic orderings, only the AFC solutions are obtained on the 2$\times$6 cluster. 
On the 3$\times$4 cluster, both the spiral and AFC solutions are obtained around $1.2 \lesssim t' \lesssim 1.3$, 
and spiral is energetically favored for $1.2 \lesssim t' \lesssim 1.25$, while the AFC is more stable for $t' \simeq 1.3$. 
Therefore the nature of the magnetic ordering in this region could not be determined. 
As will be stated later, the nature of the magnetic orderings in this region was not determined also in the 
VMC study.\cite{tocchio2}
For $1.3 \lesssim t' \lesssim 1.6$, the AFC is realized on both the 2$\times$6 and 3$\times$4 clusters. 
We remark here that we considered only the two magnetic orderings 120$^\circ$ spiral and AFC, 
therefore we can not exclude the possibility that magnetic orderings very different from these two, 
e.g. incommensurate spiral orderings, are realized in the nonmagnetic phase in Fig.~\ref{fig:phase-diagram}. 

\begin{figure}
\includegraphics[width=0.44\textwidth,bb= 165 56 494 546]{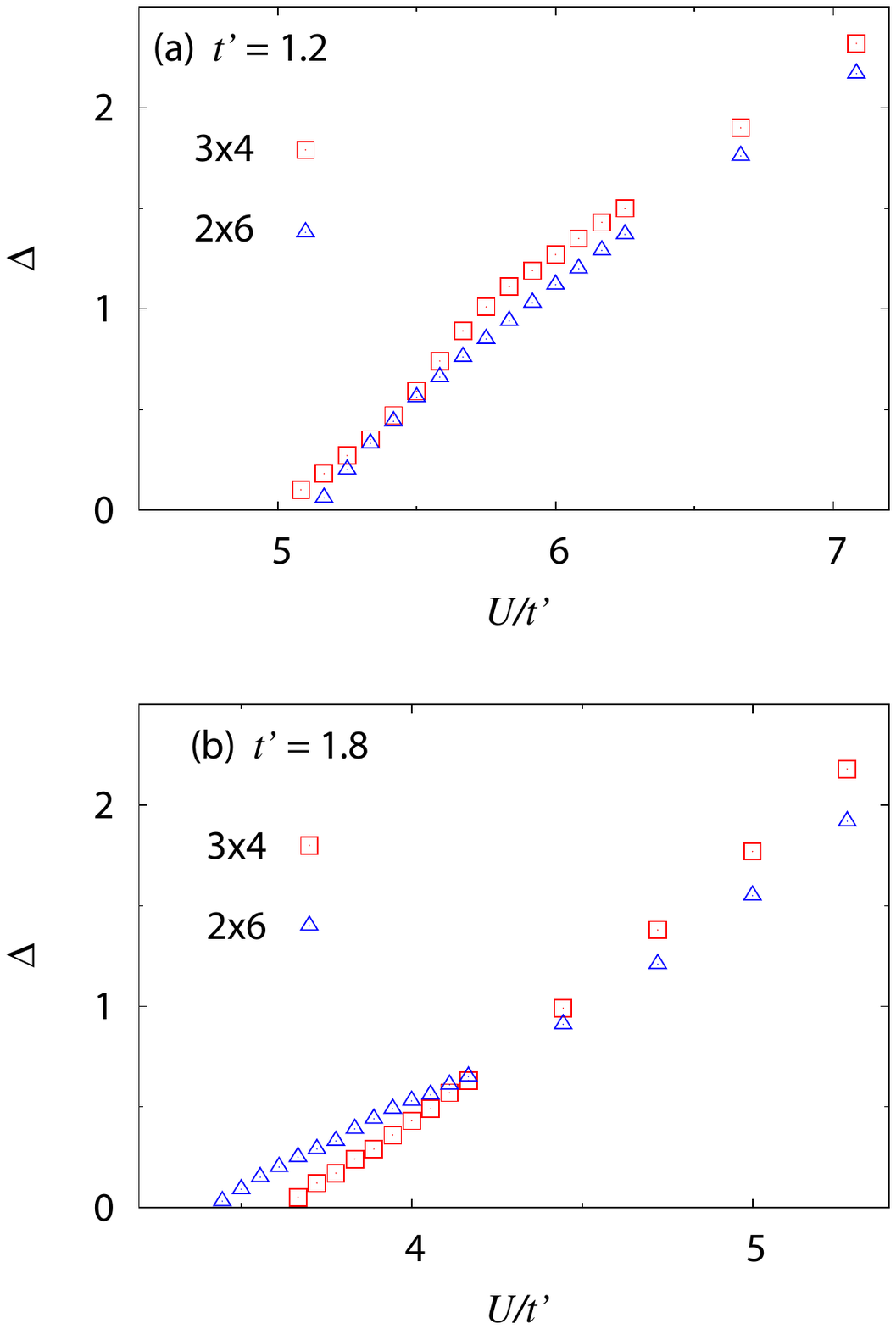}\\
\caption{
(Color online)
The Mott gaps calculated as functions of $U/t'$ by VCA assuming that no magnetic order is allowed 
on 2$\times$6 (triangles) and 3$\times$4 (squares) clusters at (a) $t' = 1.2$ and (b) and $t' = 1.8$. 
\label{fig:gap}\\[-2.2em]
}
\end{figure}

\begin{figure}
\includegraphics[width=0.47\textwidth,bb= 148 33 482 527]{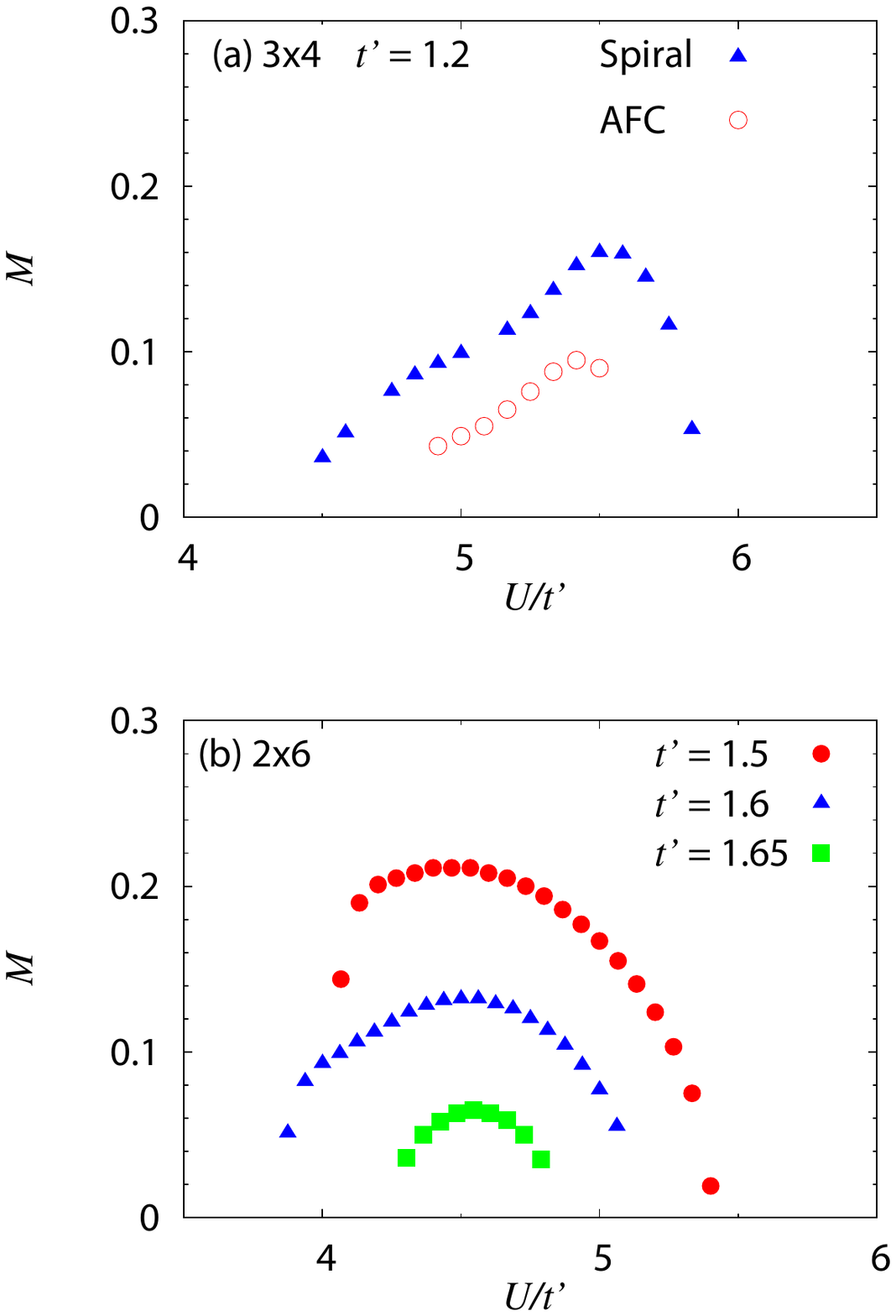}\\
\caption{
(Color online)
(a) 
The magnetic order parameters $M$ calculated on the 
3$\times$4 cluster for spiral (filled triangles) and AFC (unfilled squares) 
at $t'=1.2$ as functions of $U/t'$. AFC solutions are energetically disfavored compared to the spiral. 
(b) 
The magnetic order parameters $M$ as functions of $U/t'$ calculated on the 2$\times$6 cluster for AFC 
at $t'=1.5$ (circles), $t'=1.6$ (triangles), and $t'=1.65$ (squares). 
\label{fig:order-para}\\[-2.2em]
}
\end{figure}

We further study the Mott gap and magnetic order parameters. Fig.~\ref{fig:gap} shows the Mott gap 
calculated as a function of $U/t'$ by VCA on the 2$\times$6 (triangles) and 3$\times$4 (squares) clusters at 
(a) $t' = 1.2$ and (b) and $t' = 1.8$ assuming that no magnetic order is allowed. 
The Mott gap closes continuously at the transition point and its 
cluster shape dependence is rather small. In VCA the Mott gap closes continuously also for other lattices.\cite{yamada2011}
Fig.~\ref{fig:order-para} shows the magnetic order parameter $M$ as a function of $U/t'$ calculated (a) on the 
3$\times$4 cluster for spiral (filled triangles) and AFC (unfilled squares) at $t'=1.2$ and 
(b) on the 2$\times$6 cluster for AFC at $t'=1.5$ (circles), $t'=1.6$ (triangles), and $t'=1.65$ (squares). 
In (a) AFC solutions are energetically disfavored compared to spiral. As was stated, we could not determine 
the nature of the magnetic orderings around $1.2 \lesssim t' \lesssim 1.3$ because the cluster shape dependence is large. 
For more one dimensional region $1.3 \lesssim t' \lesssim 1.6$, only the AFC solutions are obtained 
both on the 2$\times$6 and 3$\times$4 clusters. 

Next we discuss the implications of our results on the compounds Cs$_2$CuBr$_4$ and Cs$_2$CuCl$_4$. 
As for Cs$_2$CuCl$_4$, which exhibits the spin liquid behavior,\cite{coldea2001} 
$ t'/t \simeq 1.8$ is suggested by the comparison between the neutron scattering experiments and 
theoretical calculations,\cite{coldea2001} density-functional calculations,\cite{foyevtsova2011} 
temperature dependence of the magnetic susceptibility,\cite{Zheng2005} 
and electron spin resonance spectroscopy,\cite{zvyagin2014} which implies that the $U/t'$ dependence of the insulating gap 
is given in Fig.~\ref{fig:gap} (b) and $ 3.5 \lesssim U/t'$. 
As for the Cs$_2$CuBr$_4$, the electron spin resonance spectroscopy\cite{zvyagin2014} 
reported the rather large value $ t'/t \simeq 1.5$, while the 
other studies\cite{coldea2001,Zheng2005,foyevtsova2011} mentioned above suggest $ t'/t \simeq 1.2$. 
Taking into account the fact that Cs$_2$CuBr$_4$ is a magnet with 
the spiral order, our analysis, together with the previous VCA studies\cite{yamada2014} 
implies that $ t'/t \lesssim 1.2 \sim 1.3$ at most since the spiral order is stable only 
around the isotropic point $ t'/t = 1$. 

Next we compare our results with the recent analysis of the same model by the VMC.\cite{tocchio2}
For $1.2 \lesssim t' \lesssim 1.6$ their and our results are qualitatively consistent since 
both studies predict that nonmagnetic insulator is realized for large $U$ and it changes to magnetic states 
as $U$ decreases. As for the nature of this magnetic state, 
it seems hard to determine the magnetic orderings near $t' \simeq 1.2$ not only in our study but also in their study. 
In their study more general pitch angles were analyzed in addition to 120$^\circ$ spiral. 
For $1.4 \lesssim t' \lesssim 1.6$, both their and our studies predict the AFC ordering. 
For more one-dimensional region $1.6 \lesssim t'$ their and our results are qualitatively different. 
They have studied up to $ t' \sim 3.3$ and always obtained the AFC phase between the nonmagnetic insulator and 
paramagnetic metal for whole the range, 
while in our analysis the nonmagnetic insulator becomes energetically favored compared to the AFC 
above the paramagnetic metal for $1.6 \lesssim t'$, 
thus purely paramagnetic metal-insulator transition takes place in our phase diagram. 
Quantitatively, transition point between the nonmagnetic insulator and the magnetic state obtained by VMC\cite{tocchio2} 
is $U/t' \simeq 8 \sim 12$ for $1.2 \lesssim t' \lesssim 1.6$ and is larger compared to our 
value $U/t' \simeq 5 \sim 6$, and the transition point from the magnetic state to the paramagnetic metal is 
$U/t' \simeq 4 \sim 5$ for $1.2 \lesssim t' \lesssim 1.6$ in both studies. 
Both the VMC\cite{tocchio2} and present VCA studies predict that the magnetically disordered state is 
favored compared to AFC or spiral in the Heisenberg limit for $1.2 \lesssim t'$. 
 As for the Heisenberg model, for example the existence of one-dimensional spin liquid phase is predicted 
for wide range of the interchain hopping in Ref.~\onlinecite{heisenberg1}, 
while nonmagnetic insulator is not found for $1.0 \leq t'/t $ in Ref.~\onlinecite{heisenberg2}. 
So the issue on the existence of the spin liquid phase in the region $1.0 \leq t'/t $ is not fully settled in the 
Heisenberg model.

\section{Summary and conclusions}

In summary we have studied the magnetic properties and Mott transition in the Hubbard model on the anisotropic triangular 
lattice by VCA in the region of the weakly coupled chains, and the phase diagram is analyzed at zero temperature 
and half-filling taking into account the spiral (of 120$^\circ$ pitch angle) and AFC orderings. 
We found that when the on-site Coulomb repulsion $U$ is relatively large $5 \sim 6 \lesssim U/t'$, 
nonmagnetic insulator, which is a candidate of spin liquid, is realized in the region $1.2 \lesssim t'/t \lesssim 2.0$. 
As $U$ decreases, this nonmagnetic insulator changes to a magnetic state for $1.2 \lesssim t'/t \lesssim 1.6$, and 
it changes to the paramagnetic metal for $1.6 \lesssim t'/t \lesssim 2.0$. 
Thus purely paramagnetic metal insulator transition takes place in this region. 

In our analysis, the two magnetic orderings spiral (of 120$^\circ$ pitch angle) and AFC are considered to 
investigate non-magnetic states, and we can not exclude the possibility that magnetic orderings not approximated well 
by these orderings, e.g. incommensurate spiral orderings, are realized in the nonmagnetic phase found in our study.

\section*{ACKNOWLEDGMENT}

The author would like to thank R.~Eder, T.~Inakura, H.~Kurasawa, H.~Nakada, T.~Ohama, Y.~Ohta, and K.~Seki 
for useful discussions. Parts of numerical calculations were done using the computer facilities of 
the IMIT at Chiba University and Yukawa Institute.


\begin{thebibliography}{99}

\bibitem{lefebvre00}
S. Lefebvre, P. Wzietek, S. Brown, C. Bourbonnais, D. J{\'e}rome, C. M{\'e}zi{\`e}re, 
M. Fourmigu{\'e}, and P. Batail, 
Phys. Rev. Lett. {\bf 85}, 5420 (2000).

\bibitem{shimizu03}
Y. Shimizu, 
K. Miyagawa, K. Kanoda, M. Maesato, and G. Saito, 
Phys. Rev. Lett. {\bf 91}, 107001 (2003).

\bibitem{kanoda3}
F. Kagawa, T. Itou, K. Miyagawa, and K. Kanoda, Phys. Rev. B {\bf 69}, 064511 (2004). 

\bibitem{manna} R.S. Manna, M. de Souza, A. Br\"uhl, J.A. Schlueter, and M. Lang, {\prl} {\bf 104}, 016403 (2010).

\bibitem{cong2011} P. T. Cong, B. Wolf, M. de Souza, N. Krueger, A.A. Haghighirad, 
   S. Gottlieb-Schoenmeyer, F. Ritter, W. Assmus, I. Opahle, K. Foyevtsova, H.O. Jeschke, 
   R. Valenti, L. Wiehl, and M. Lang, \prb {\bf 83}, 064425 (2011).
   (2013).
\bibitem{ono2004} T. Ono, H. Tanaka, O. Kolomiyets, H. Mitamura, T. Goto, K. Nakajima, A. Oosawa, 
   Y. Koike, K. Kakurai, J. Klenke, P. Smeibidle and M. Meissner, J. Phys.: Condens. Matter 
   {\bf 16}, S773 (2004).
\bibitem{coldea2001} R. Coldea, D.A. Tennant, A.M. Tsvelik, and Z. Tylczynski, 
   \prl {\bf 86}, 1335 (2001); R. Coldea, D.A. Tennant, and Z. Tylczynski, 
   \prb {\bf 68}, 134424 (2003).
\bibitem{foyevtsova2011} K. Foyevtsova, I. Opahle, Y.-Z. Zhang, H.O. Jeschke, R. Valenti,
   \prb {\bf 83}, 125126 (2011).
\bibitem{ono2005} T. Ono, H. Tanaka, T. Nakagomi, O. Kolomiyets, H. Mitamura, F. Ishikawa, 
   T. Goto, K. Nakajima, A. Oosawa, Y. Koike, K. Kakurai, J. Klenke, P. Smeibidle, M. Meissner, 
   and H. Aruga Katori, J. Phys. Soc. Jpn. {\bf 74}, Suppl.  135 (2005).
\bibitem{Zheng2005} W. Zheng, R. R. P. Singh, R. H. McKenzie, and R. Coldea, 
     \prb{\bf 71}, 134422 (2005). 
\bibitem{zvyagin2014} S.A. Zvyagin, D. Kamenskyi, M. Ozerov, J. Wosnitza, M. Ikeda, T. Fujita, 
   M. Hagiwara, A.I. Smirnov, T.A. Soldatov, A.Ya. Shapiro, J. Krzystek, R. Hu, H. Ryu, 
   C. Petrovic, and M.E. Zhitomirsky, \prl {\bf 112}, 077206 (2014).



\bibitem{Senechal00}
D.~S\'en\'echal, D. Perez, and M. Pioro-Ladri\'ere, Phys. Rev. Lett. {\bf 84}, 522 (2000); 
D.~S\'en\'echal, D. Perez, and D. Plouffe, Phys. Rev. B {\bf 66}, 075129 (2002).

\bibitem{Potthoff:2003-1}
M.~Potthoff, M.~Aichhorn, and C.~Dahnken, Phys. Rev. Lett. {\bf 91} 206402 (2003); 
C.~Dahnken, M.~Aichhorn, W. Hanke, E. Arrigoni, and M.~Potthoff, Phys. Rev. B {\bf 70}, 245110 (2004).

\bibitem{Potthoff:2003} M. Potthoff, Eur. Phys. J. B \textbf{32}, 429 (2003).

\bibitem{tocchio2}
L.F. Tocchio, C. Gros, R. Valent\'i, F. Becca, \prb {\bf 89}, 235107 (2014).

\bibitem{lw}
L. M. Luttinger and J. C. Ward, 
Phys. Rev. {\bf 118}, 1417 (1960). 


\bibitem{aichhorn}   
M. Aichhorn, E. Arrigoni, M. Potthoff, and W. Hanke, 
Phys. Rev. B {\bf 74}, 024508 (2006). 

\bibitem{yamada2011} A. Yamada, K. Seki, R. Eder, and Y. Ohta, Phys. Rev. B {\bf 83}, 195127 (2011); 
Phys. Rev. B {\bf 88}, 075114 (2013).
\bibitem{yamada2014} A. Yamada, \prb {\bf 89}, 195108 (2014).
\bibitem{heisenberg1}  
S.~Yunoki and S.~Sorella, Phys. Rev. B {\bf 74}, 014408 (2006); 
M.~Q.~Weng, D. N. Sheng, Z. Y. Weng, and R. J. Bursill, Phys. Rev. B {\bf 74}, 012407 (2006); 
Y. Hayashi and M. Ogata, J. Phys. Soc. Jpn. {\bf 76}, 053705 (2007); 
D. Heidarian, S.~Sorella and F. Becca, Phys. Rev. B {\bf 80}, 012404 (2009); 
J. Reuther and R. Thomale, Phys. Rev. B {\bf 83}, 024402 (2011). 
\bibitem{heisenberg2}  
ZhengWeihong, R.~H.~McKenzie, R.~P.~Singh, Phys. Rev. B {\bf 59}, 14367 (1999); 
O.~A.~Starykh and L.~Balents, {\prl} {\bf 98}, 077205 (2007);
S.~Ghamari, C.~Kallin, S.~-S.~Lee, and E.~S.~S{\o}rensen, Phys. Rev. B {\bf 84}, 174415 (2011)
A.~Weichselbaum and S.~R.~White, Phys. Rev. B {\bf 84}, 245130 (2011). 



\end{thebibliography}
\end{document}